\documentclass[12pt]{article}
\pdfoutput=1
\usepackage{amsmath,amssymb,amsfonts}
\usepackage{a4wide,epsfig,psfrag,scalefnt}
\usepackage[dvipsnames]{xcolor}
\usepackage{braket}
\usepackage{placeins}
\usepackage{breqn}
\usepackage{slashed}
\usepackage{enumitem}
\usepackage[numbers,sort&compress]{natbib}
\usepackage{caption}
\usepackage{subcaption}

\usepackage{fancyvrb}

\graphicspath{ {figs/} }

\allowdisplaybreaks

\begin{document}

\title{\vskip-3cm{\baselineskip14pt
    \begin{flushleft}
     \normalsize P3H-23-043, TTP23-024, ZU-TH 34/23
    \end{flushleft}} \vskip1.5cm
  Towards $gg\to HH$ at next-to-next-to-leading order: light-fermionic 
  three-loop corrections 
  }

\author{
  Joshua Davies$^{a}$,
  Kay Sch\"onwald$^{b}$,
  Matthias Steinhauser$^{c}$
  \\
  {\small\it (a) Department of Physics and Astronomy, University of Sussex,
    Brighton BN1 9QH, UK}
  \\
  {\small\it (b) Physik-Institut, Universit\"at Z\"urich, Winterthurerstrasse 190,}\\
  {\small\it 8057 Z\"urich, Switzerland}
  \\
  {\small\it (c) Institut f{\"u}r Theoretische Teilchenphysik,
    Karlsruhe Institute of Technology (KIT),}\\
  {\small\it Wolfgang-Gaede Stra\ss{}e 1, 76128 Karlsruhe, Germany}
}

\date{}

\maketitle

\thispagestyle{empty}

\begin{abstract}

\noindent
We consider light-fermion three-loop corrections to $gg\to HH$ using forward
scattering kinematics in the limit of a vanishing Higgs boson mass, which
covers a large part of the physical phase space. We compute
the form factors and discuss the technical challenges. The approach outlined
in this letter can be used to obtain the full virtual corrections to
$gg\to HH$ at next-to-next-to-leading order.

\end{abstract}


\newpage


\section{\label{sec::intro}Introduction}

The simultaneous production of two Higgs bosons is a promising process to
obtain information about their self-coupling in the scalar sector of the
Standard Model and beyond. Its study will be of primary importance after the
high-luminosity upgrade of the Large Hadron Collider and thus it is important
that there are precise predictions from the theory side.

The cross section for Higgs boson pair production is dominated by the gluon-fusion
process, which is loop-induced~\cite{Glover:1987nx}. Thus, at
next-to-leading (NLO) order the virtual corrections require the computation of
two-loop four-point function with massive internal top quarks. There are
numerical results which take into account the full dependence of all mass
scales~\cite{Borowka:2016ehy,Borowka:2016ypz,Baglio:2018lrj}. Furthermore,
there are a number of analytic approximations which are valid in various
limits, which cover different parts of the phase space.  Particularly appealing approaches
have been presented in Refs.~\cite{Bonciani:2018omm,Bellafronte:2022jmo,Davies:2023vmj} where
the expansion around the forward-scattering kinematics has been combined with
the high-energy expansion and it has been shown that the full phase space can
be covered. Thus, these results are attractive alternatives to computationally
expensive purely numerical approaches.

Beyond NLO, current results are based on expansions for large top quark
masses. Results in the infinite-mass limit are available at
NNLO~\cite{deFlorian:2013jea,deFlorian:2013uza,Grigo:2014jma} and
N$^3$LO~\cite{Chen:2019lzz,Chen:2019fhs} and finite $1/m_t$ corrections have
been considered at NNLO in Refs.~\cite{Grigo:2015dia,Davies:2019djw,Davies:2021kex}.
 
In Ref.~\cite{Baglio:2020wgt} the renormalization scheme dependence on the top
quark mass has been identified as a major source of uncertainty of the NLO
predictions.  In general, such uncertainties are reduced after including
higher-order corrections, i.e., virtual
corrections at NNLO including the exact dependence on the top quark mass. This
requires the computation of $2\to 2$ scattering amplitudes at three-loop order
with massive internal quarks; this is a highly non-trivial problem. Current
analytic and numerical methods are not sufficient to obtain results with full
dependence on all kinematic variables, as is already the case at two loops.
However, after an expansion in the
Mandelstam variable $t$ (see
Refs.~\cite{Bellafronte:2022jmo,Degrassi:2022mro,Davies:2023vmj}) and the
application of the ``expand and match''~\cite{Fael:2021kyg,Fael:2021xdp,Fael:2022miw}
method to compute the master integrals, one
obtains semi-analytic results which cover a large part of the phase space.
Such a result allows the study of the renormalizations scheme dependence at
three-loop order. In this letter we outline a path to the three-loop
calculation and present first results for the light-fermionic corrections.

Let us briefly introduce the kinematic variables describing the $2\to 2$ process, with
massless momenta $q_1$ and $q_2$ in the initial state and massive momenta
$q_3$ and $q_4$ in the final state.  It is convenient to introduce the
Mandelstam variables as
\begin{eqnarray}
  s = (q_1+q_2)^2\,,\qquad t = (q_1+q_3)^2\,,\qquad u = (q_1+q_4)^2\,,
\end{eqnarray}
where all momenta are incoming. For $gg \to HH$ we have
\begin{eqnarray}
  q_1^2=q_2^2=0\,,\qquad q_3^2=m_H^2\,,\qquad q_4^2=m_H^2\,,
\end{eqnarray}
and the transverse momentum of the final-state particles is given by
\begin{eqnarray}
  p_T^2 &=&\frac{u\,t-m_H^4}{s}\,.
            \label{eq::pT}
\end{eqnarray}

For Higgs boson pair production one can identify two linearly independent
Lorentz structures
\begin{eqnarray}
  A_1^{\mu\nu} &=& g^{\mu\nu} - {\frac{1}{q_{12}}q_1^\nu q_2^\mu
  }\,,\nonumber\\
  A_2^{\mu\nu} &=& g^{\mu\nu}
                   + \frac{1}{{p_T^2} q_{12}}\left(
                   q_{33}    q_1^\nu q_2^\mu
                   - 2q_{23} q_1^\nu q_3^\mu
                   - 2q_{13} q_3^\nu q_2^\mu
                   + 2q_{12} q_3^\mu q_3^\nu \right)\,,
\end{eqnarray}
where $q_{ij} = q_i\cdot q_j$, which allows us to introduce two
form factors in the amplitude
\begin{eqnarray}
  {\cal M}^{ab} &=& 
  \varepsilon_{1,\mu}\varepsilon_{2,\nu}
  {\cal M}^{\mu\nu,ab}
  \,\,=\,\,
  \varepsilon_{1,\mu}\varepsilon_{2,\nu}
  \delta^{ab} X_0 s 
  \left( F_1 A_1^{\mu\nu} + F_2 A_2^{\mu\nu} \right)
  \,.
                    \label{eq::M}
\end{eqnarray}
Here $a$ and $b$ are adjoint colour indices
and $X_0 = G_F/2\sqrt{2} \times T_F \alpha_s(\mu)/(2\pi)$
with $T_F=1/2$. $G_F$ is Fermi's constant and $\alpha_s(\mu)$ is the strong
coupling constant evaluated at the renormalization scale $\mu$.
We write the perturbative expansion of the form factors
as
\begin{eqnarray}
  F &=& F^{(0)} + \left(\frac{\alpha_s(\mu)}{\pi}\right) F^{(1)}
        + \left(\frac{\alpha_s(\mu)}{\pi}\right)^2 F^{(2)} 
        + \cdots
  \,,
  \label{eq::F}
\end{eqnarray}
and decompose $F_1$ and $F_2$ into ``triangle'' and ``box'' form
factors
\begin{eqnarray}
  F_1^{(k)} &=& \frac{3 m_H^2}{s-m_H^2} F^{(k)}_{\rm tri}+F^{(k)}_{\rm box1}
                \,, \nonumber\\
  F_2^{(k)} &=& F^{(k)}_{\rm box2}\,.
                \label{eq::F_12}
\end{eqnarray}
In this notation $F^{(k)}_{\rm box1}$ and $F^{(k)}_{\rm box2}$ contain both
one-particle irreducible and reducible contributions. The latter
appear for the first time at two-loop order; exact results for
the so-called ``double-triangle'' contributions can be found in~\cite{Degrassi:2016vss}.

Analytic results for the leading-order form factors are available
from~\cite{Glover:1987nx,Plehn:1996wb} and the two-loop triangle form factor
has been computed in
Refs.~\cite{Harlander:2005rq,Anastasiou:2006hc,Aglietti:2006tp}. The main
focus of this letter is on the light-fermionic contribution to the three-loop quantities
$F^{(2)}_{\rm box1}$ and $F^{(2)}_{\rm box2}$ for $t=0$ and $m_H=0$.
Expansions around the large top quark mass limit of $F^{(2)}_{\rm tri}$,
$F^{(2)}_{\rm box1}$ and $F^{(2)}_{\rm box2}$ can be found in
Ref.~\cite{Davies:2019djw} and results for $F^{(2)}_{\rm tri}$ valid for all
$s/m_t^2$ have been computed in
Refs.~\cite{Davies:2019nhm,Davies:2019roy,Harlander:2019ioe,Czakon:2020vql}.

\begin{figure}
  \centering
  \includegraphics[width=0.32\textwidth]{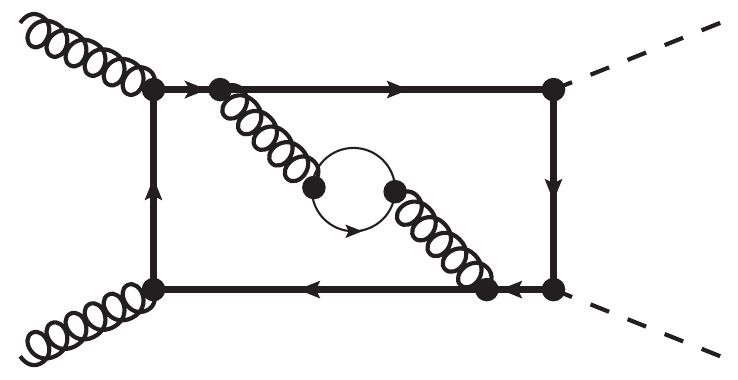}
  \includegraphics[width=0.32\textwidth]{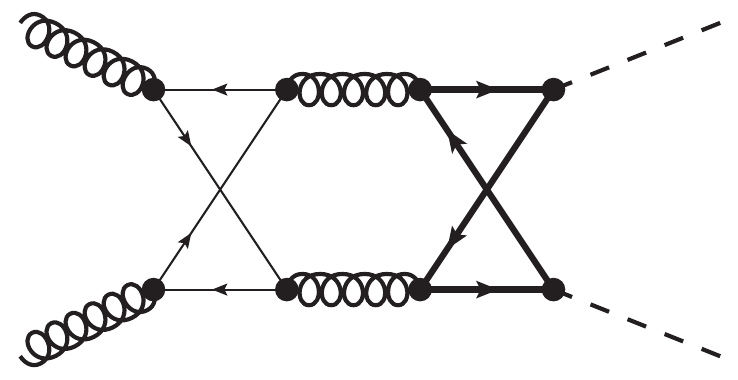}
  \includegraphics[width=0.32\textwidth]{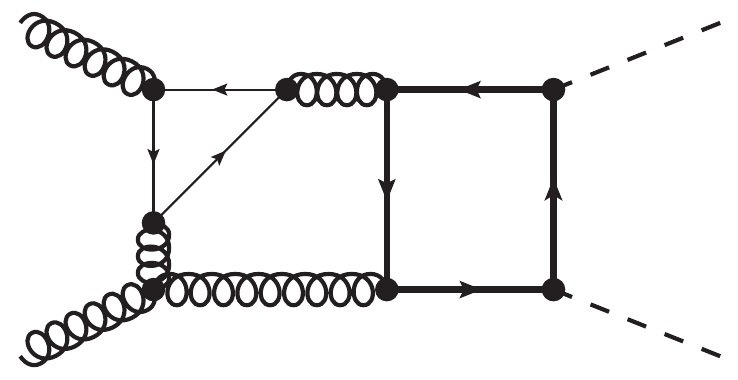}
  \includegraphics[width=0.32\textwidth]{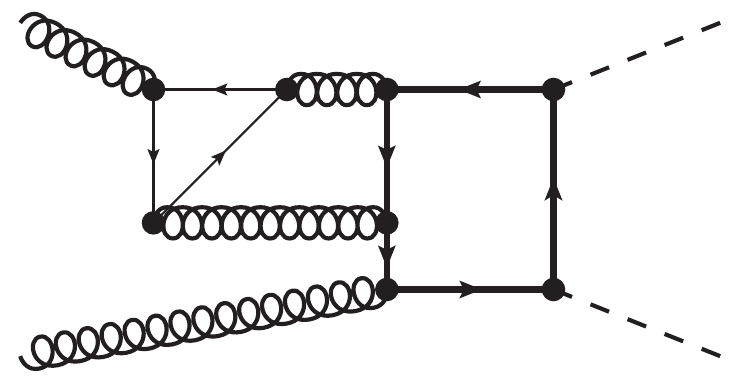}
  \includegraphics[width=0.32\textwidth]{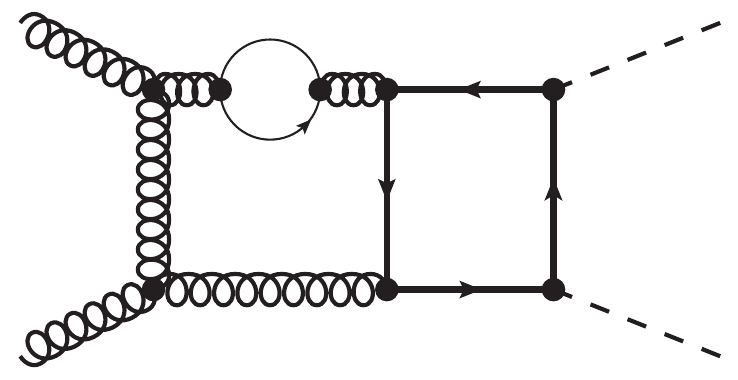}
  \caption{Sample Feynman diagrams. Curly lines denote gluons, dashed ones Higgs bosons, while 
  thin (thick) lines are massless (top) quarks.}
  \label{fig::diags}
\end{figure}

We decompose the three-loop form factors as
\begin{eqnarray}
  F^{(2)} &=& n_lT_F F^{(2),n_l}
  \,\,=\,\, n_lT_F \left(C_F F^{FL} + C_A F^{AL}\right) + \ldots
              \,,
              \label{eq::F3loop}
\end{eqnarray}
where the ellipses stand for further colour factors which we do not consider here.
Sample Feynman diagrams contributing to $F^{FL}$ and $F^{AL}$
are shown in Fig.~\ref{fig::diags}.

In this letter we consider $t=0$ and $m_H=0$, i.e.~the leading term in an
expansion around $t\to 0$ and $m_H\to 0$. This constitutes a crude approximation,
however, in a large part of the phase space it contributes a
major part of the corrections.
For example, choosing $t=0$ and $m_H=0$ at two loops (NLO),
at a transverse momentum of $p_T=100$~GeV the form factor $F_{\rm box1}$
deviates from its exact value by at most
30\%, depending on the value of $\sqrt{s}$ considered. This means that
more than two thirds of the form factor value are covered by the $t=0$, $m_H=0$
approximation. Furthermore,
we concentrate on the one-particle irreducible contributions.
We note that $F_{\rm box2}$ vanishes for $t=0$.
More details are given below in Section~\ref{sec::gghh}.

We present here results for the light-fermionic (``$n_l$'') terms and show that this
approach can be used to obtain the three-loop virtual corrections to $gg\to HH$. The
remaining contributions contain many more integral topologies and more complicated
integrals, which have to be integration-by-parts (IBP) reduced to master integrals.

In the next section we outline the techniques used for the calculations
and discuss the results in Section~\ref{sec::gghh}.
In Section~\ref{sec:conclusions} we conclude and provide an outlook
for the computation of the full corrections.


\section{\label{sec::tech}Technical Details}

The basic philosophy of our calculation has already been outlined in
Ref.~\cite{Davies:2023vmj}, where the two-loop amplitude for $gg\to HH$ has
been considered in the small-$t$ and high-energy limit and it has been shown
that the combination of both expansions covers the whole phase-space. The
starting point for both expansions is the amplitude expressed in terms of
the same master integrals which are obtained from a reduction problem which involves
the dimensional variables $s$, $t$ and $m_t$.\footnote{A Taylor expansion in
  $m_H$ in a first step eliminates the Higgs boson mass from the reduction
  problem.}  Using currently available tools such a reduction is not possible
at three loops.
To avoid such an IBP reduction, one can try to expand the unreduced amplitude in the
respective limit.  The high-energy expansion is obtained via a complicated
asymptotic expansion which involves a large number of different regions. On
the other hand, the limit $t\to 0$ leads to a simple Taylor expansion which
can be easily realized at the level of the integrands.
Furthermore, the expansion around forward-scattering kinematics covers
a large part of the physically relevant phase space~\cite{Bellafronte:2022jmo}.

Our computation begins by generating the amplitude with {\tt
  qgraf}~\cite{Nogueira:1991ex}, and then using {\tt
  tapir}~\cite{Gerlach:2022qnc} and {\tt
  exp}~\cite{Harlander:1997zb,Seidensticker:1999bb} to map the diagrams onto
integral topologies and convert the output to {\tt FORM}~\cite{Ruijl:2017dtg}
notation. The diagrams are then computed with the in-house ``{\tt calc}''
setup, to produce an amplitude in terms of scalar Feynman integrals. These
tools work together to provide a high degree of automation.  We perform
the calculation for general QCD gauge parameter which drops out once the
amplitude is expressed in terms of master integrals. This is a welcome check
for our calculation.

The scalar integrals can be Taylor expanded in $m_H$ at this point, as done at
two loops in Refs.~\cite{Davies:2018ood,Davies:2018qvx,Davies:2023vmj},
however at three loops in this letter we keep only the leading term in this
expansion, i.e., set $m_H=0$.

The next step is to expand the amplitude around the forward kinematics
($t\to 0$) at the integrand level. This is implemented in {\tt FORM} by
introducing $q_\delta = q_1+q_3$ in the propagators and expanding in
$q_\delta$ to the required order. Note that $q_\delta^2=t$. After treating
the tensor integrals, where $q_\delta$ appears contracted with a loop momentum,
we need to perform a partial-fraction decomposition to eliminate linearly
dependent propagators. The partial fractioning rules are produced automatically
by {\tt tapir} when run with the forward kinematics ($q_3=-q_1$)
specified\footnote{In an alternative approach, we have also used
{\tt LIMIT}~\cite{Herren:2020ccq} to generate the partial fractioning rules.}.
Note that although for the present publication we compute the ``$t=0$ contribution'',
we must properly expand in $q_\delta$ to produce the amplitude to order $t^0$
due to inverse powers of $t$ appearing in the projectors. These
inverse powers ultimately cancel in the final result.
This procedure yields amplitudes
for $F_{\rm box1}$ and $F_{\rm box2}$ in terms of scalar Feynman integrals
which belong to topologies which depend only on $s$ and $m_t$ (and not on $t$).

At this point the amplitudes are written in terms of 60 integral
topologies, however these are not all independent; they can be reduced to a smaller
set by making use of loop-momentum shifts and identification of common sub-sectors.
In one approach we find these rules with the help of {\tt LiteRed}~\cite{Lee:2013mka},
which identifies a minimal set of 28 topologies.
In a second approach we use {\tt Feynson}~\cite{Magerya:2022esf} to generate these maps
and end up with 53 topologies.
The difference in the number of topologies is due to {\tt LiteRed} mapping
topology sub-sectors, while  we used {\tt Feynson} only at the top level.
When considering the full amplitude, i.e., not just the light-fermionic corrections,
only the {\tt Feynson} approach is feasible for performance reasons. It is also
possible to use {\tt Feynson} to find sub-sector mappings, which we will also use
when considering the full amplitude (which is written initially in terms of
522 integral topologies).

The amplitude is now ready for a reduction to master integrals using
{\tt Kira}~\cite{Klappert:2020nbg} and
\texttt{FireFly}~\cite{Klappert:2020aqs,Klappert:2019emp}.
The most complicated integral topology took about a week on a 16-core node,
using around 500GB of memory. After minimizing the final set of master integrals
across the topologies with {\tt Kira}, we are left with 177 master integrals
to compute. Comparing results obtained via the {\tt LiteRed} and
{\tt Feynson} topology-mapping approaches reveals one additional relation within
this set which is missed by {\tt Kira}, however, we compute the set of 177
master integrals which was first identified.

To compute the master integrals, we first establish a system of differential
equations w.r.t.~$x=s/m_t^2$. Boundary conditions are provided in the
large-$m_t$ ($x\to 0$) limit: we prepare the three-loop integrals in the
forward kinematics, and pass them to {\tt exp} which automates the asymptotic
expansion in the limit that $m_t^2 \gg s$. This leads to three-loop vacuum
integrals, as well as products of one- and two-loop vacuum integrals with
two- and one-loop massless $s$-channel $2\to 1$ integrals, respectively.
This expansion leads to tensor vacuum integrals, which our ``{\tt calc}'' setup can
compute up to rank 10. We compute the first two expansion terms in $s/m_t^2$ for
each of the 177 master integrals. To fix the boundary constants for the differential
equations we only need about half of the computed coefficients; the
rest serve as consistency checks.

The differential equations are then used to produce 100 expansion terms for
the forward-kinematics master integrals in the large-$m_t$ limit which we use
to compute $F_{\rm box1}$. Since these results are analytic in the large-$m_t$
limit we can compare with the results obtained in Ref.~\cite{Davies:2019djw}
in the limit $t=0$, and find agreement.

The final step is to use the ``expand and match'' approach~\cite{Fael:2021kyg,Fael:2021xdp,Fael:2022miw}
to obtain ``semi-analytic'' results which cover the whole $s$ range. Note that
this approach properly takes into account the threshold effects at the point
$s = 4m_t^2$. ``Semi analytic'' means that our final results consist of expansions
around a set of $x$ values, where the expansion coefficients are available
only numerically. Starting from the (analytic) expansion around $x=0$, each
expansion provides numeric boundary conditions to fix the coefficients of the
subsequent expansion. Each expansion is only ever evaluated within its radius
of convergence.


\section{\label{sec::gghh}Three-loop light-fermionic contributions to $F_{\rm box1}$}

\begin{figure}[t]
  \centering
  \mbox{
    \includegraphics[width=0.45\textwidth]{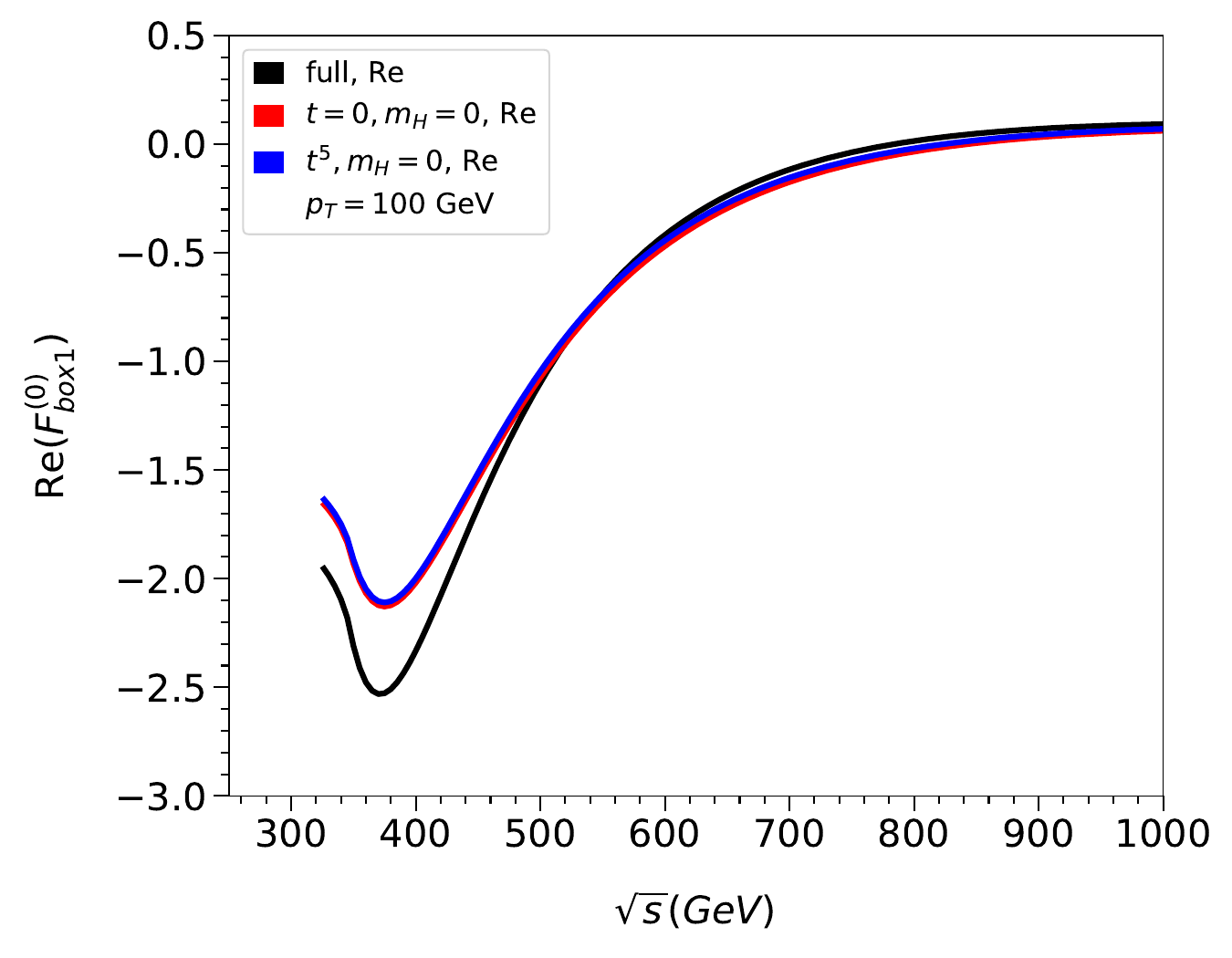}
    \includegraphics[width=0.45\textwidth]{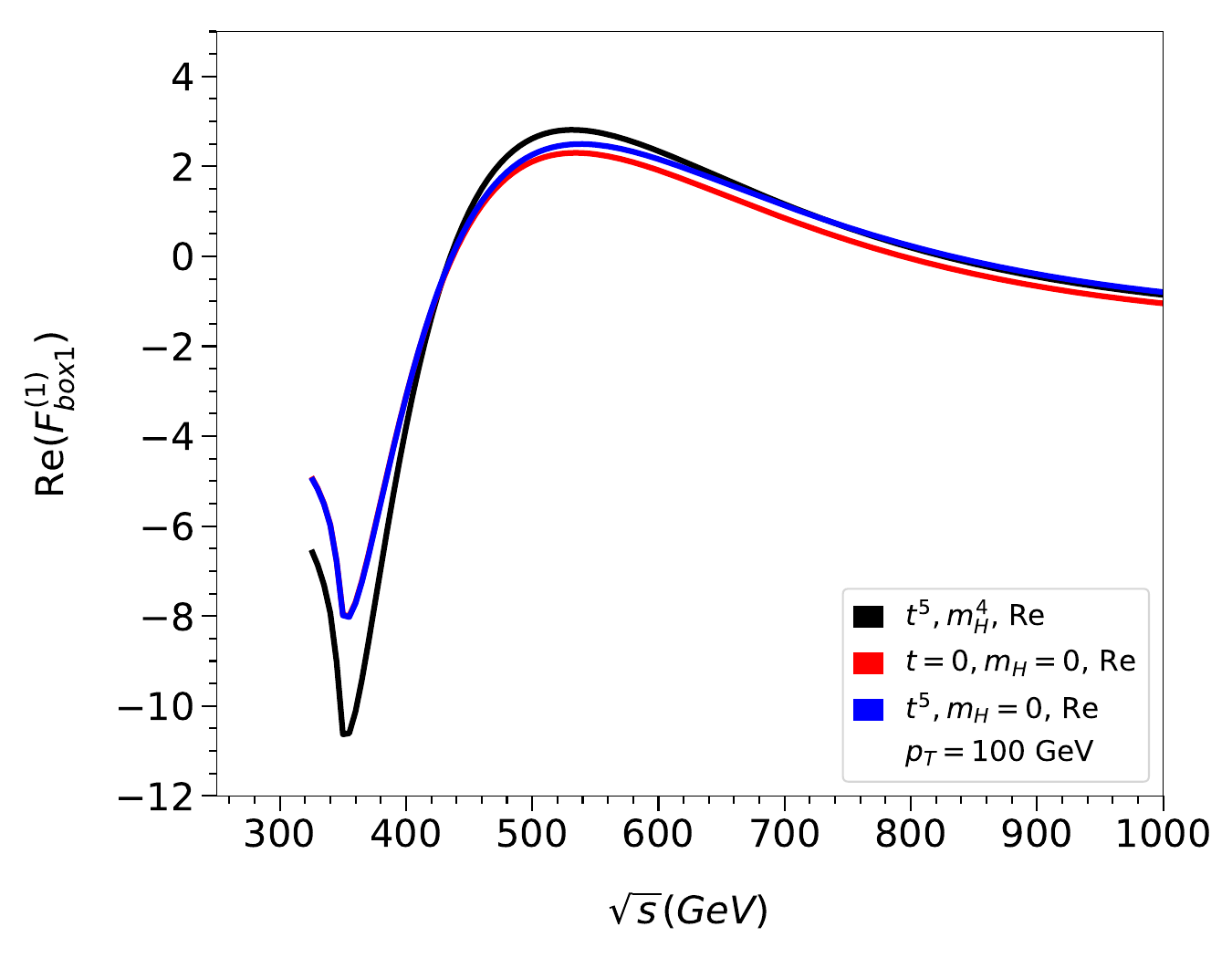}
  }
  \caption{The real part of $F_{\rm box1}$ at one and two loops, for
    $p_T = 100$~GeV.  The $t=0$, $m_H=0$ approximation is shown in red, and
    the $t^5, m_H=0$ approximation in blue.  At one loop we compare with the
    exact result with full $m_H$ and $t$ dependence, in black.  At two loops,
    in lieu of an exact result, we compare with the $t^5, m_H^4$
    approximation, in black.}
  \label{fig::F1_1l2l_100}
\end{figure}

\begin{figure}[t]
  \centering
  \mbox{
    \includegraphics[width=0.3\textwidth]{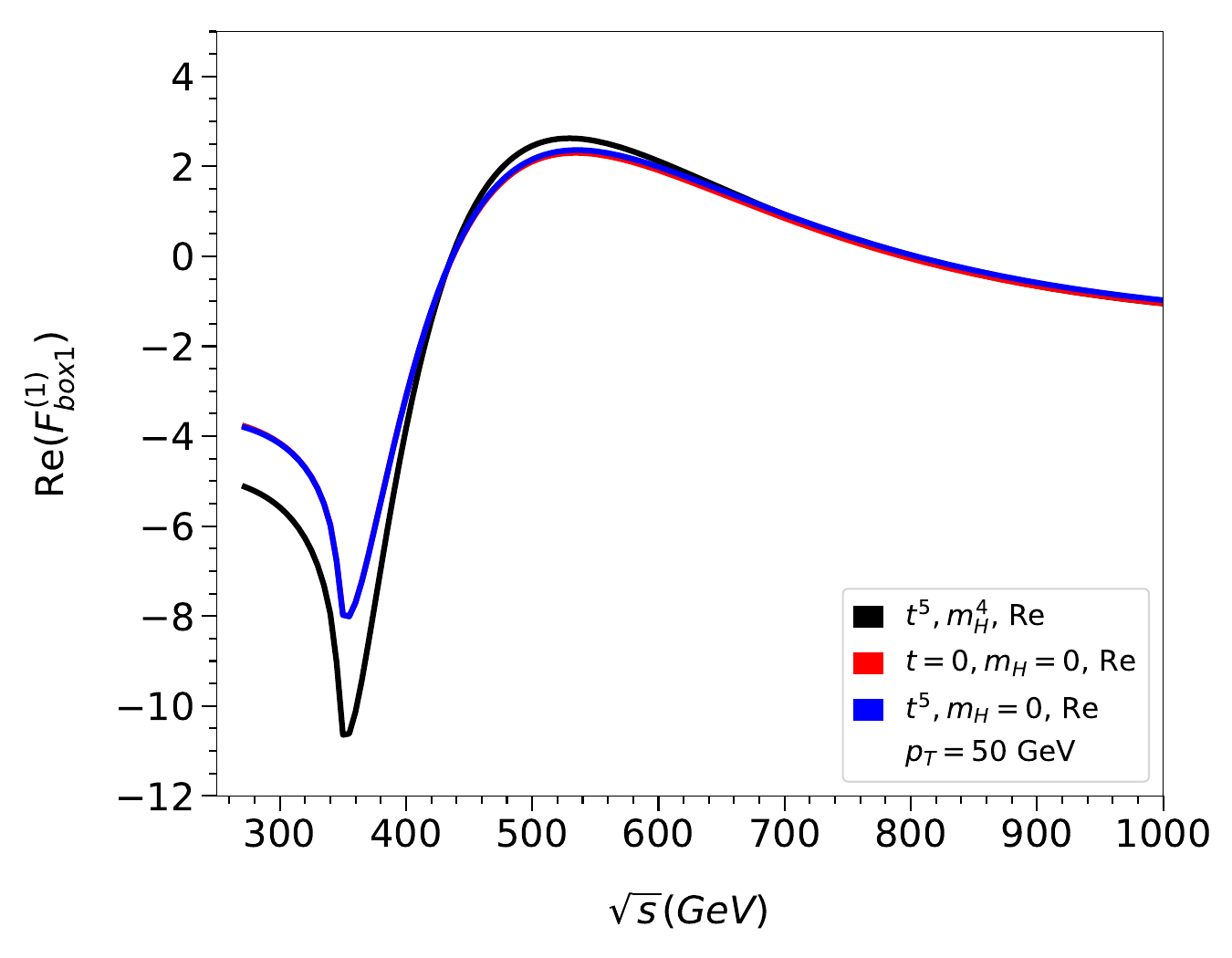}
    \includegraphics[width=0.3\textwidth]{F1_2l_pT100_t0mhs0.pdf}
    \includegraphics[width=0.3\textwidth]{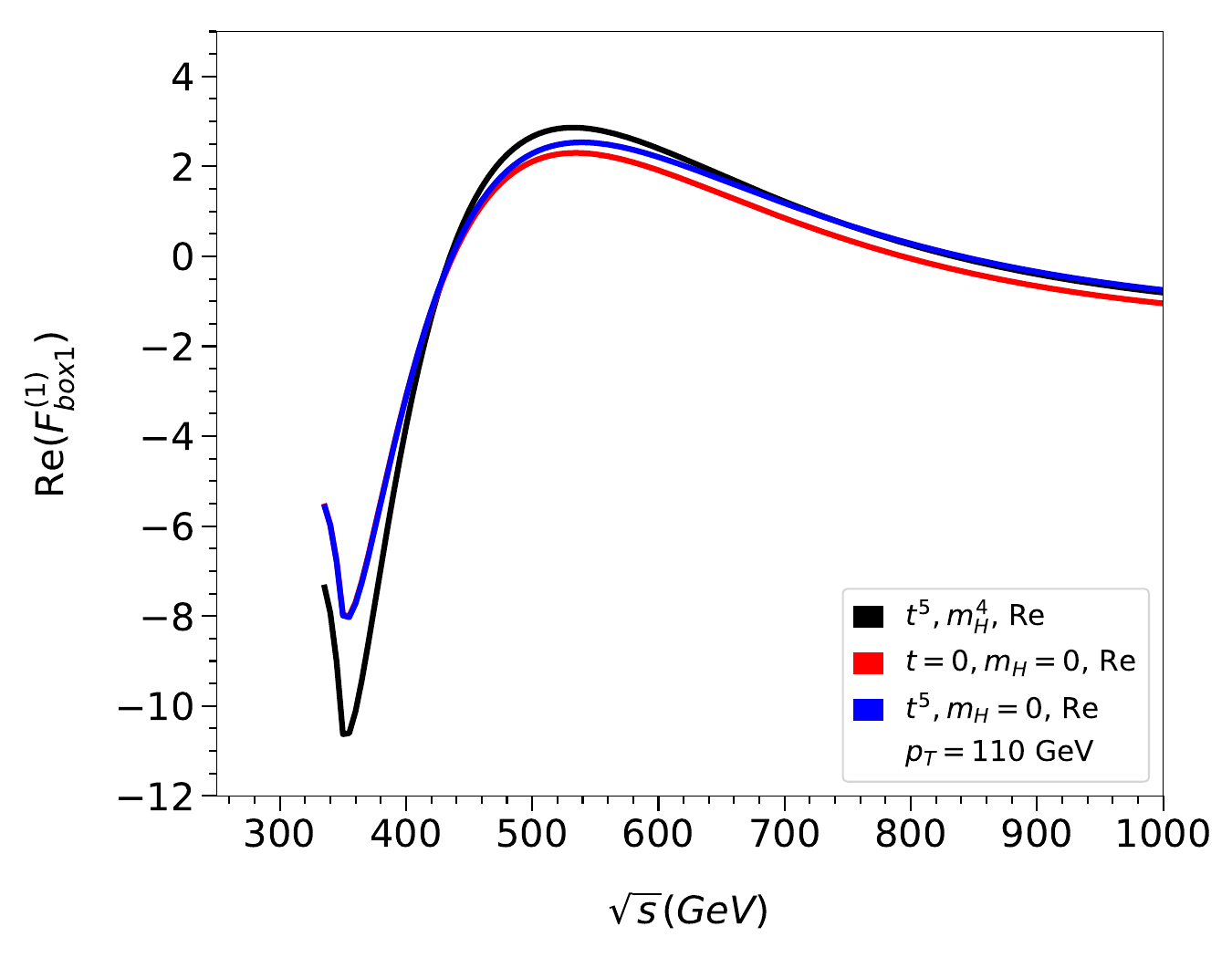}
  }
  \\
  \mbox{
    \includegraphics[width=0.3\textwidth]{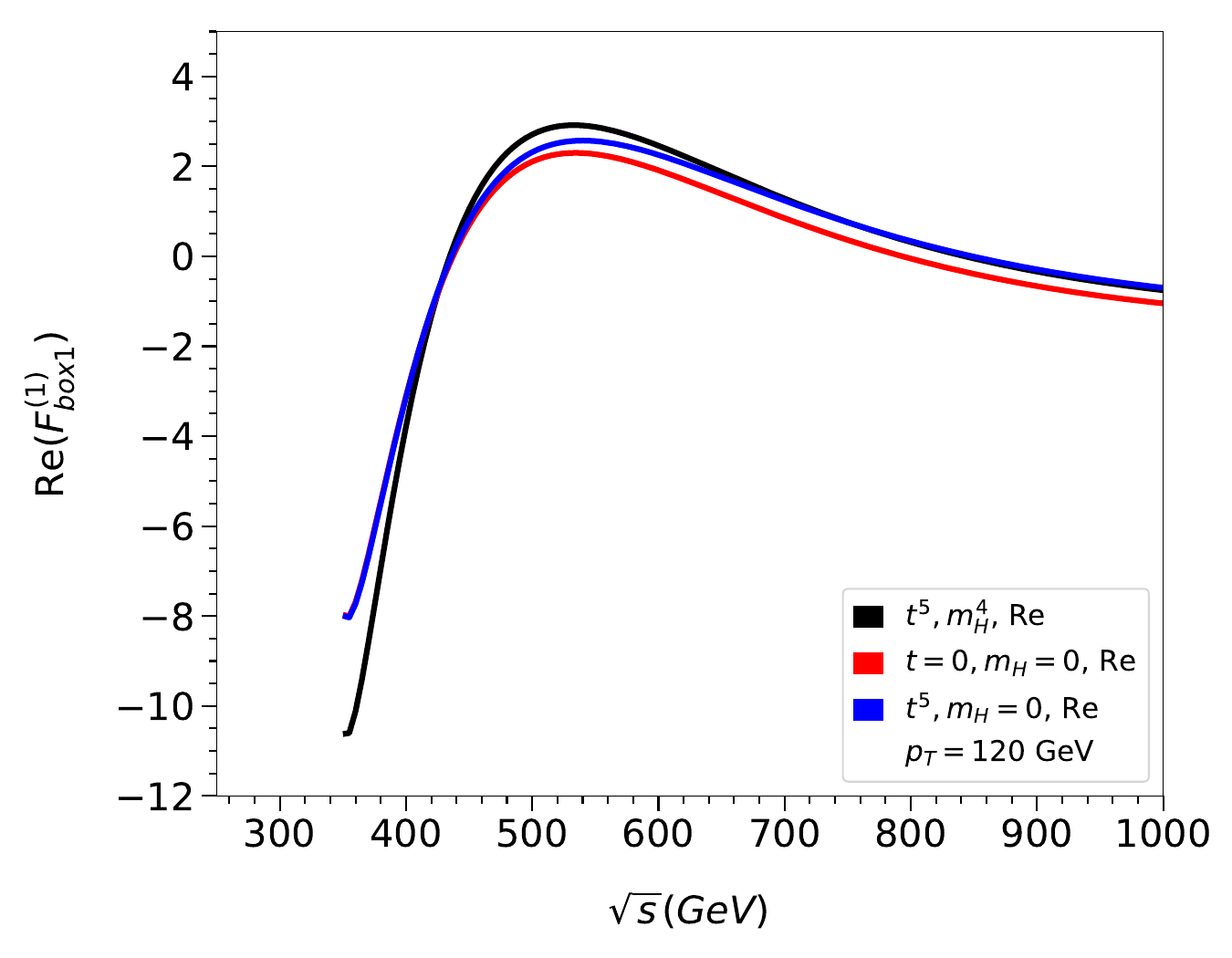}
    \includegraphics[width=0.3\textwidth]{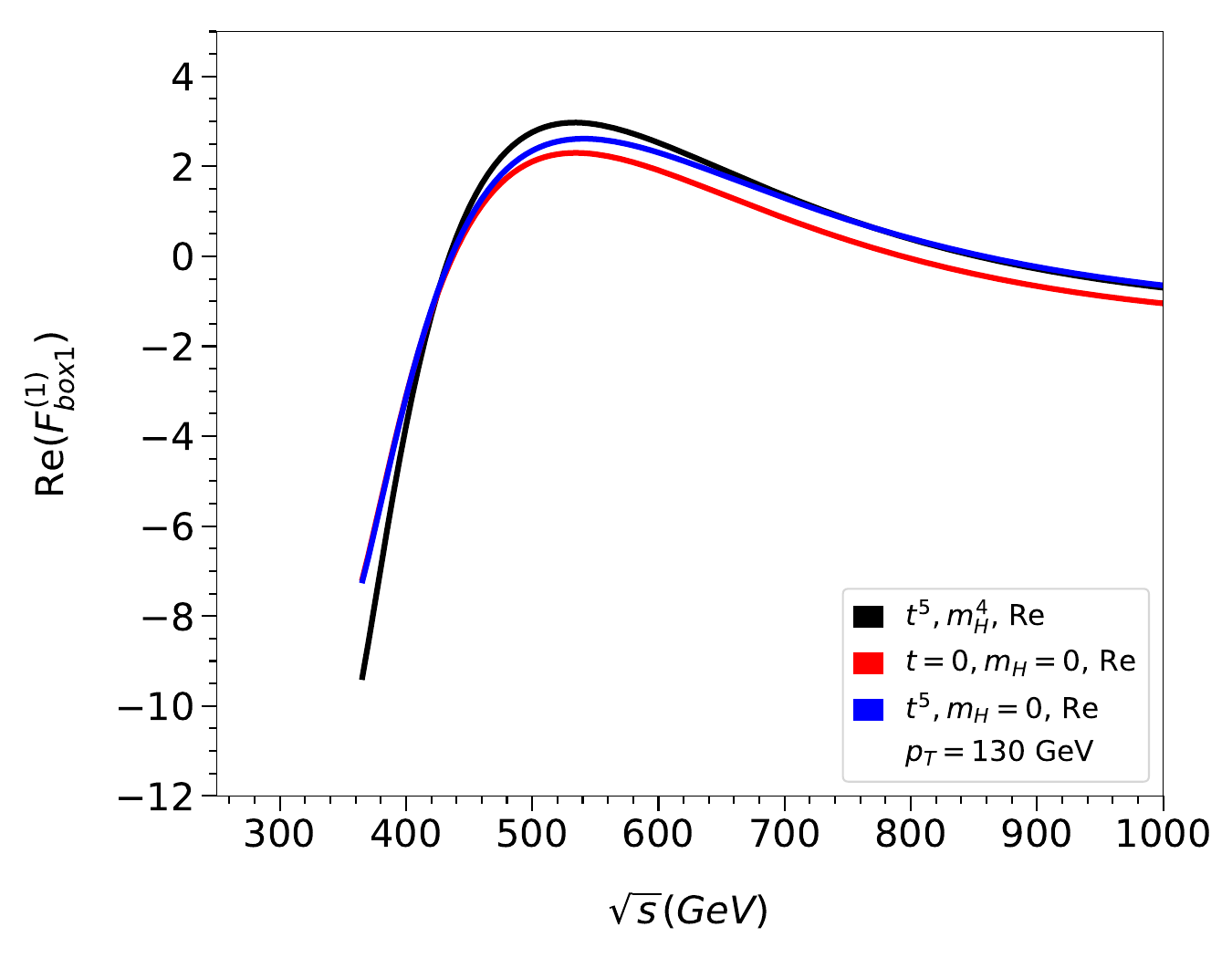}
    \includegraphics[width=0.3\textwidth]{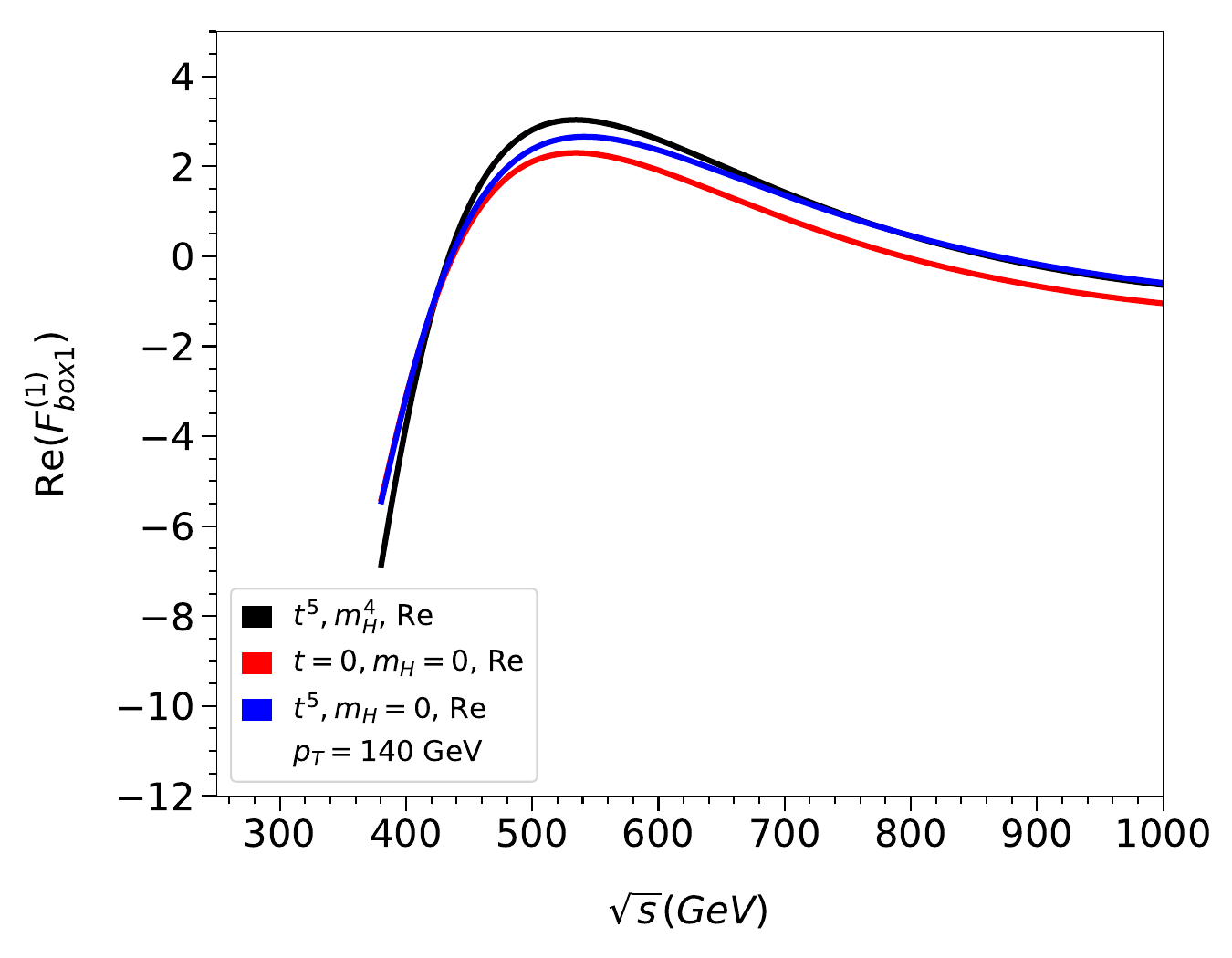}
  }
  \\
  \mbox{
    \includegraphics[width=0.3\textwidth]{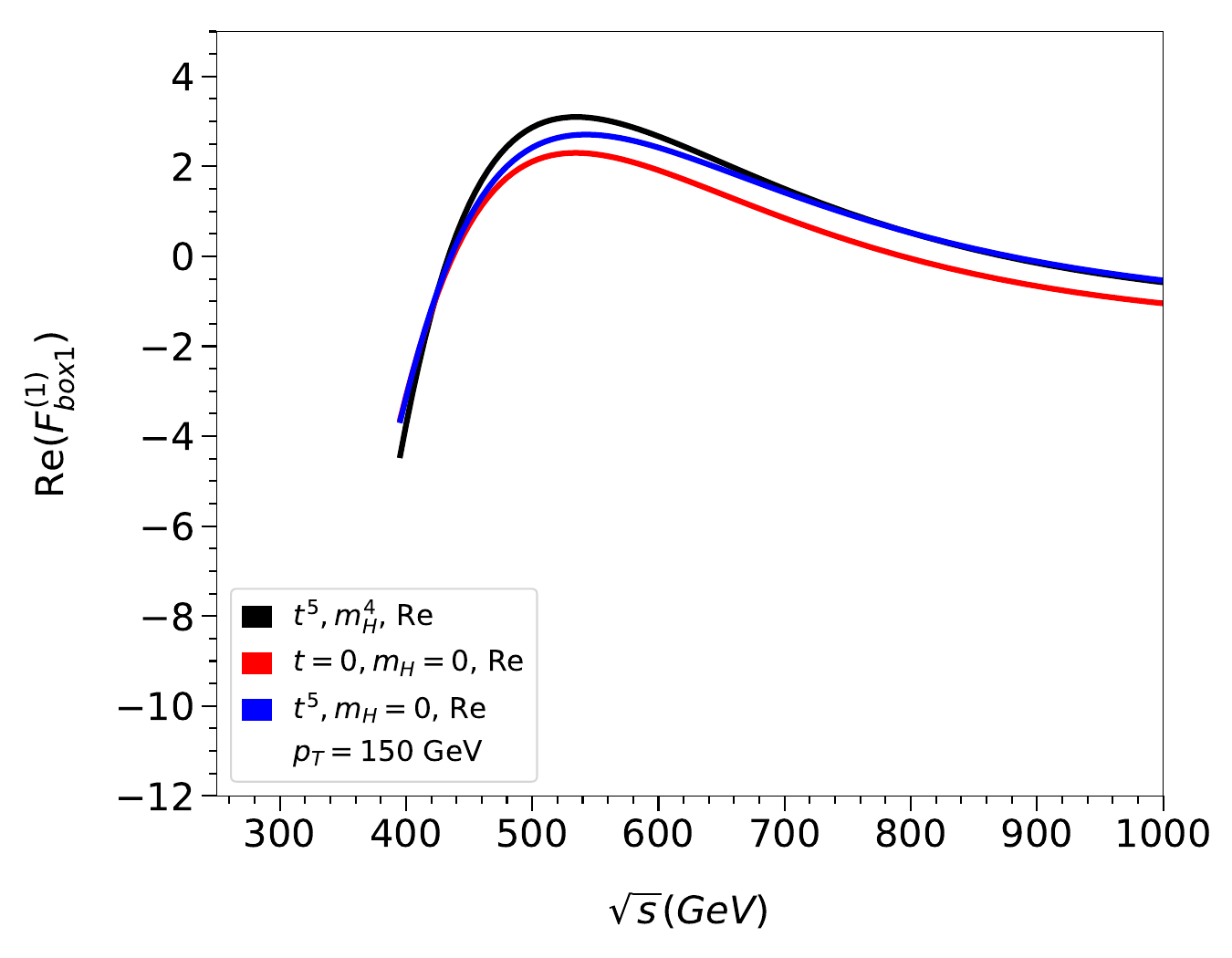}
    \includegraphics[width=0.3\textwidth]{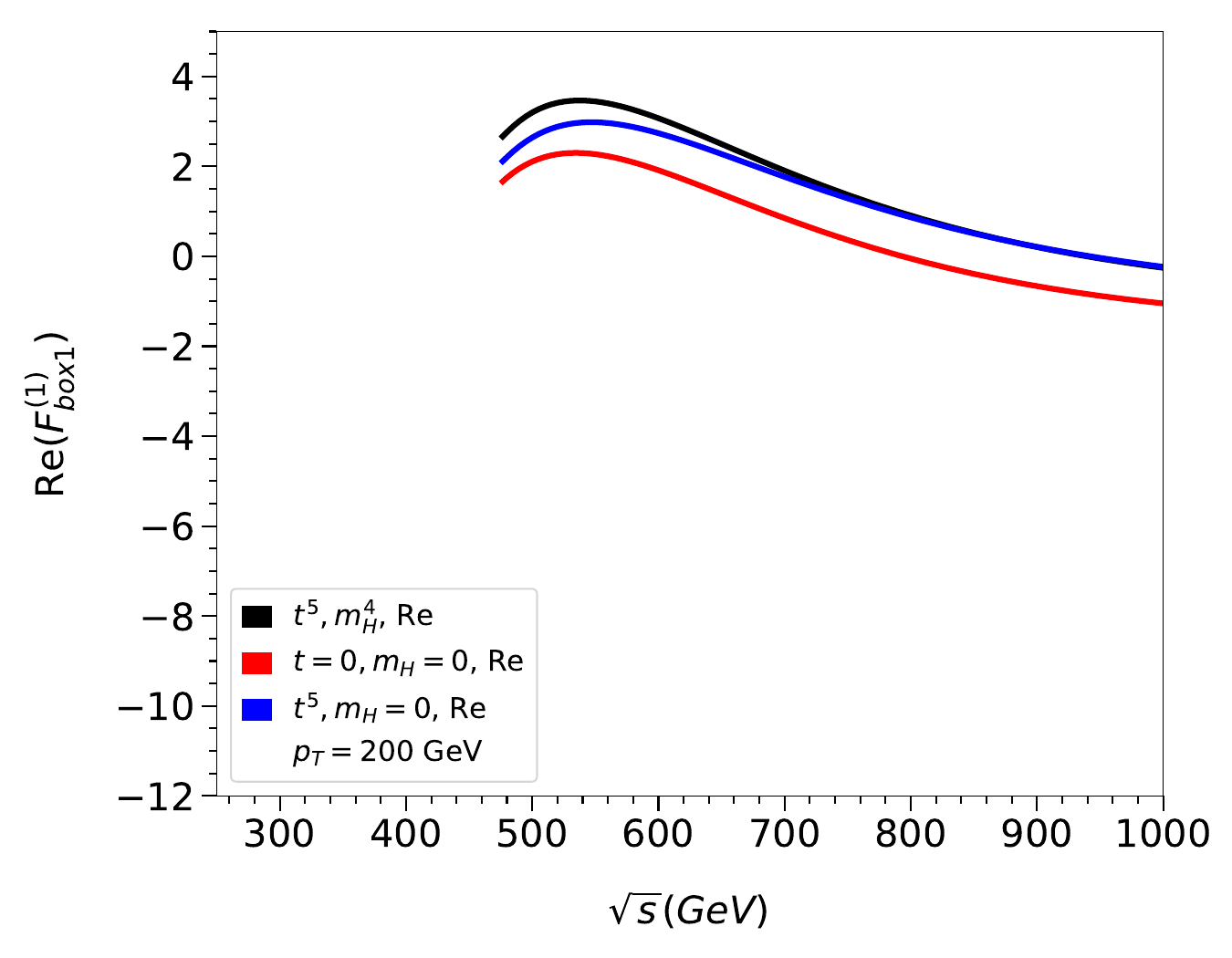}
  }
  \caption{Two-loop results for various values of $p_T$. The meaning of the
    curves is described in the caption to Fig.~\ref{fig::F1_1l2l_100}.
    Note that the red curves are independent of $p_T$ and thus they are
    identical in all panels.}
  \label{fig::F1_2l}
\end{figure}

\begin{figure}[t]
  \centering
  \includegraphics[width=0.45\textwidth]{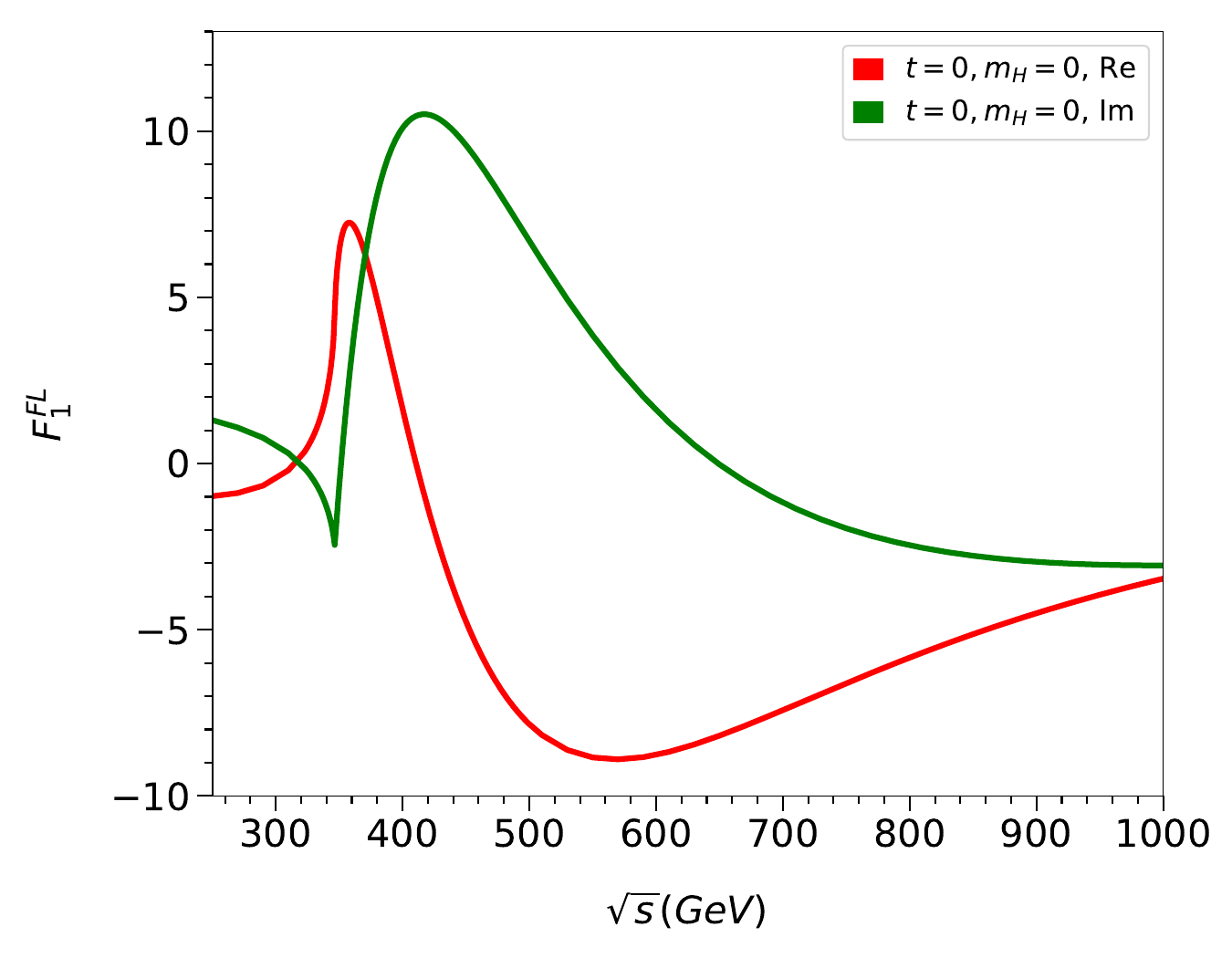}
  \includegraphics[width=0.45\textwidth]{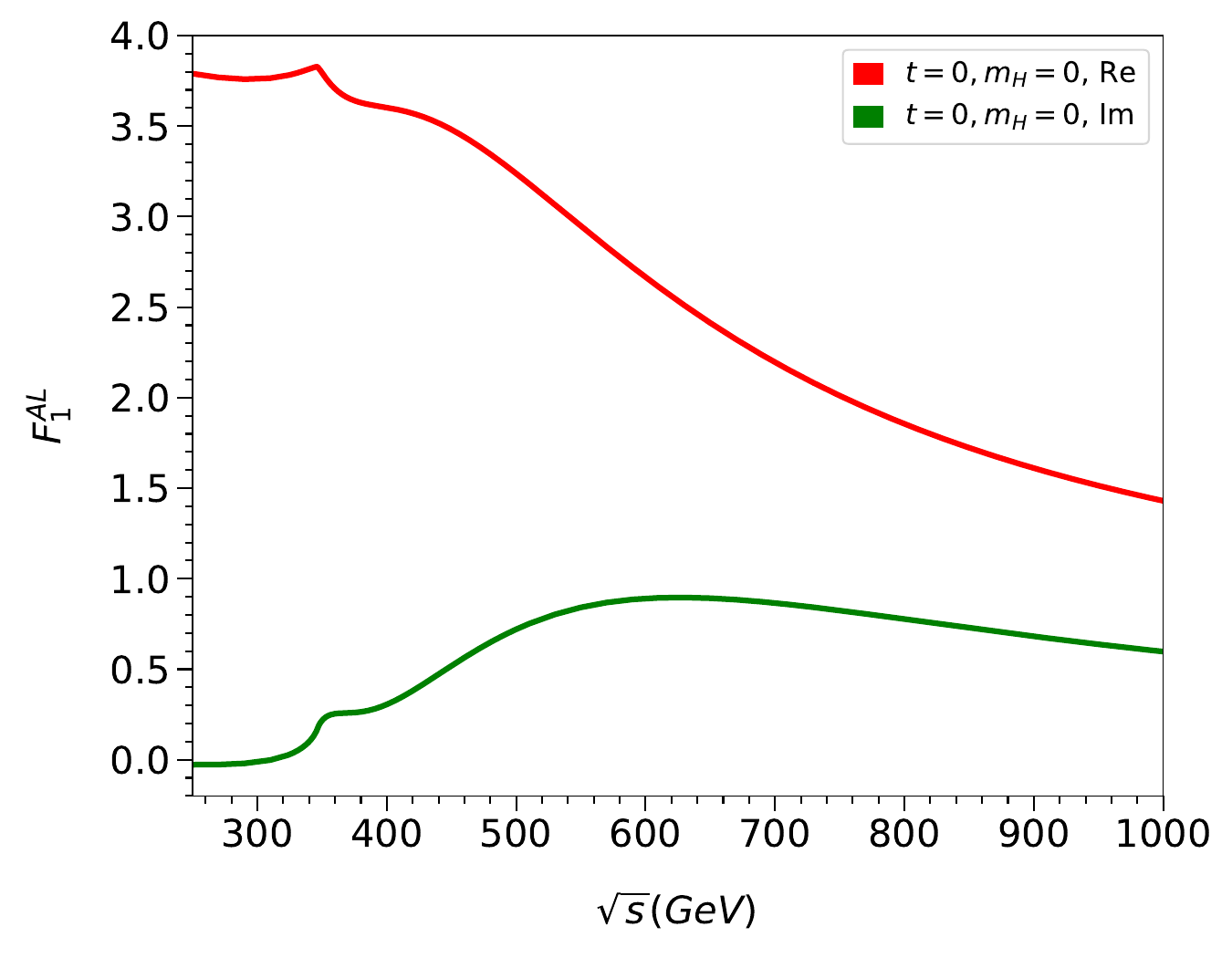}
  \\
  \includegraphics[width=0.45\textwidth]{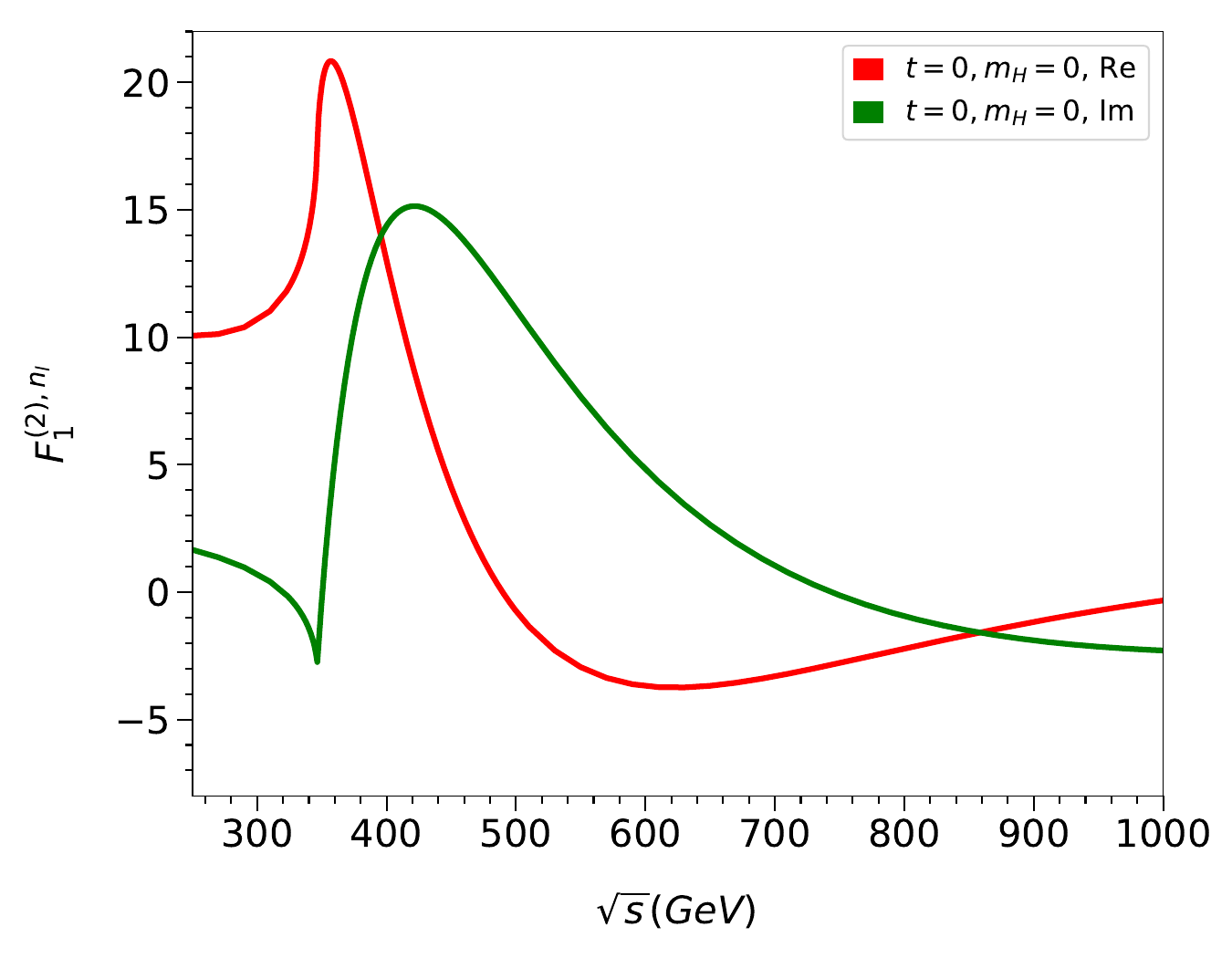}
  \caption{Real (red) and imaginary (green) parts of $F_{\rm box1}^{FL}$, $F_{\rm
      box1}^{AL}$ and $F_{\rm box1}^{(2),n_l}$
    as a function of $\sqrt{s}$.}
  \label{fig::F12_3nl}
\end{figure}

In this section we present the light-fermionic three-loop corrections to
the form factor $F_{\rm box1}$ for Higgs boson pair production. We note again
that in our $t=0, m_H=0$ approximation, $F_{\rm box2}$ vanishes; we observe this
after IBP reduction and writing the result in terms of the minimal set of master
integrals.

We obtain the renormalized form factors after the renormalization of the
parameters $\alpha_s$ and $m_t$ and the wave functions of the gluons in the
initial state. We then express our results in terms of $\alpha_s^{(5)}$ and
treat the remaining infrared divergences following
Ref.~\cite{Catani:1998bh}.\footnote{For more details see Section~4 of
  Ref.~\cite{Davies:2019djw} where analytic large-$m_t$ results for $F_{\rm box1}$ and
  $F_{\rm box2}$ have been computed at three-loop order.}  This leads to finite results
for $F_{\rm box1}$.
In the following we present numerical results.  For the top quark and Higgs
boson masses, we use the values $m_t = 173.21$~GeV and $m_H=125.1$~GeV.

Let us first discuss the one- and two-loop results.  In
Fig.~\ref{fig::F1_1l2l_100}
we show the real part of $F_{\rm box1}$ for $p_T=100$~GeV.
In red, we show the approximation that we use at three loops, i.e., $t=0$ and $m_H=0$.
In black, we show curves with the full dependence on $t$ and $m_H$. At one loop
this is the fully exact result, but at two loops this is an expansion to order
$t^5$ and $m_H^4$; we have shown in Ref.~\cite{Davies:2023vmj} that this provides
an extremely good approximation of the (unknown) fully exact result.
We observe that the $t=0$, $m_H=0$ curves approximate the ``exact'' results with an
accuracy of about 30\% in the region below about $\sqrt{s}=500$~GeV.
For higher energies the approximation works better.

In Fig.~\ref{fig::F1_1l2l_100} we also show blue curves which include
expansion terms up to $t^5$, but still only the leading term in the $m_H$
expansion. These curves lie very close to the red $t=0$, $m_H=0$ curves, which
show that for $p_T\approx 100$~GeV it is more important to incorporate
additional terms in the $m_H$ expansion than in the $t$ expansion.
For higher values of $p_T$ we expect that higher $t$ expansion terms
become more important. This can be seen in Fig.~\ref{fig::F1_2l} where results
of the two-loop form factor are shown for various values of $p_T$.
The panels also show that a large portion of the cross section
is covered by the $t=0$ approximation, even for
$p_T=200$~GeV where, for lower values of $\sqrt{s}$, about 50\% are
captured by the red curve.

In Fig.~\ref{fig::F12_3nl} we show the new results
obtained in this letter. The plots show both the real (in red) and imaginary
(in green) parts
of the light-fermionic part of $F_{\rm box1}$, both separated into the $C_F$
and $C_A$ colour factor contributions, and their combination.
The form factor $F_1^{\rm FL}$  shows  a strong variation around the
top quark pair threshold region whereas in the case
of $F_1^{\rm AL}$ only small bumps are observed. This behaviour is 
more pronounced than at two-loop order where
we observe a leading logarithmic contribution which goes like
$v \log{v}$, where $v = \sqrt{1-4m_t^2/s}$. At three loops we find an
additional power of $\log{v}$ which is responsible for the larger variation
around this point.

If we assume the same convergence pattern for the expansion in $t$ and
$m_H$ as at one- and two-loop order the results shown in
Fig.~\ref{fig::F12_3nl} approximate the (unknown) exact result for the
light-fermion contribution at the level of 30\%. This is supported by
the large-$m_t$ results where NNLO corrections to the form factors
have been computed in Ref.~\cite{Davies:2019djw}. For parameters
within the range of validity of the large-$m_t$ approximation we confirm
that the $t=0$, $m_H=0$ approximation lies within the
30\% range compared to the results in the large-$m_t$ approximation
where the full dependence on $t$ and $m_H$ is retained.

Taking into account that $n_l=5$,
the numerical value of the light-fermionic contribution to $F_{\rm box1}$ at
three-loops exceeds the size of the two-loop form factor by almost an
order of magnitude. Although this is compensated by the additional factor
of $\alpha_s/\pi$, this hints at sizeable three-loop corrections. However,
for a final conclusion, the remaining diagrams need to be computed. The
full computation will also allow a study of the top quark mass scheme
dependence. These issues will be addressed in a future publication.


\section{Conclusions}
\label{sec:conclusions}

The computation of three-loop corrections to $2\to2$ scattering processes with
massive internal particles is a technically challenging
task. Currently-available techniques are most likely not sufficient to obtain
analytic or numerical results without applying any approximation.  In this
letter we apply the ideas of
Refs.~\cite{Bonciani:2018omm,Bellafronte:2022jmo,Degrassi:2022mro,Davies:2023vmj} to $gg\to HH$
and show that three-loop corrections can be obtained.  We concentrate on the
light-fermionic three-loop contributions which is a well-defined and
gauge-invariant subset. The obtained results are valid for $t=0$ and $m_H=0$
which approximates the full result to 30\% or better for $p_T\approx 100$~GeV.

The approach outlined in this letter can also be used to compute the remaining
colour factor contributions, which are needed to study the overall impact of the
three-loop virtual corrections and also the top quark mass renormalization scheme
dependence.

In addition to the remaining colour factors, we ultimately aim to compute
the $t^1\,m_H^2$ approximation which would address the 30\% error discussed in
Section~\ref{sec::gghh}, improve the approximation for higher values of $p_T$,
and provide a non-zero value for $F_{\rm box2}$. To compute these terms will
require significantly more CPU time and, most likely, improvements to IBP reduction
software in order to efficiently reduce the large numbers of integrals produced by
the expansions.



\section*{Acknowledgements}  

We would like to thank Go Mishima for many useful discussions and Fabian
Lange for patiently answering our questions concerning {\tt Kira}.
This research was supported by the Deutsche
Forschungsgemeinschaft (DFG, German Research Foundation) under grant 396021762
--- TRR 257 ``Particle Physics Phenomenology after the Higgs Discovery''
and has received funding from the European Research Council (ERC) under
the European Union's Horizon 2020 research and innovation programme grant
agreement 101019620 (ERC Advanced Grant TOPUP).
The work of JD was supported by the Science and Technology Facilities Council (STFC) under
the Consolidated Grant ST/T00102X/1.






\end{document}